\documentclass[preprint]{sigplanconf}

\usepackage{amssymb}
\usepackage{amsmath}
\usepackage{graphicx}
\usepackage{listings}
\usepackage{capt-of}

\begin{document}



        
\title{A new Programming Paradigm with Separated Domain for Behaviour Programming and Data Management Programming}

\authorinfo{Chengpu Wang}
           {Independent Researcher}
           {Chengpu@gmail.com}

\maketitle

\begin{abstract}

A new behaviour descriptive entity type called spec is proposed, which combines the traditional interface with test rules and test cases, to completely specify the desired behaviour of each method, and to enforce the behaviour-wise correctness of all compiled units.  Using spec, a new programming paradigm is proposed, which allows the separation programming space into 1) a behaviour domain to aggregate all behaviour programming in the format of specs, 2) a object domain to bind each concrete spec to its data representation in a particular address space, and 3) a realization domain to connect the behaviour domain and the object domain. Such separation guarantees the strictness of behaviour satisfaction at compile time, while allows flexibility of dynamical binding of actual implementation at runtime.  A new convention call type expressiveness to allow data exchange between different programming languages and between different software environments is also proposed.

\end{abstract}

\category{D.3.3}{Language Constructs and Features}{Abstract data types}

\terms
Computer Programming Language, 
Programming Paradigm, 
Serialization

\keywords
Computer Programming Language, 
Programming Paradigm, 
Serialization

\section{Introduction}

The evolution of programming languages \cite{history} can be viewed as a progress on how well an object can express itself.  Assembly languages \cite{assembly} such as Intel 8086 \cite{x86} series add symbolic instructions and method call to machine language.  Procedural languages \cite{pragmatics} such as C \cite{c} add data type, API and method callback to assembly languages.  Object-oriented languages \cite{pragmatics} such as C++ \cite{c++} add dynamic cast and interface to procedural languages.  Reflective languages \cite{reflective} such as Java \cite{java} and C\# \cite{csharp} add reflection and the ability to enumerate implemented interfaces to object-oriented languages. From behaviour perspective, an object is no more than its collection of interfaces \cite{pragmatics}.  The ability to enumerate implemented interfaces in reflective languages is crucial for run-time query of objects, which is crucial for behaviour-centric programming methodology \cite{pragmatics}.  For example, recent development of semantic web \cite{semantic web} allows a piece of code to be writing from behaviour perspective rather than from intended usage perspective, and its usefulness is discovered by querying for different intended usage.  The ability of dynamic programming allows answer to behaviour requirement at runtime, to liberate the codes from prefabricated behaviours at compile time.

At this moment, the universal tool for expressiveness between software is \emph{interface} \cite{interface}, which is a what-to-call contract between a client and a service provider.  This paper focuses on explicit interface, which is usually in the format of a set of method signatures. Currently, there are three major problems when interfaces are used to specify behaviours:
\begin{enumerate}

\item Interface itself can only specify little desired behaviour in code, while relying heavily on documentation for its functionality.  For example, an interface for addition can only specify the possible result type and the possible operand types in addition to a function name.  The disparity between the desired behaviour and the actual behaviour is usually a major source of software bugs.

\item The actual behaviour of an interface usually depends on how the interface is implemented.  For example, because all the primitive integer types has limited bit count, due to value overflow, adding two positive integer operands may results in an integer value smaller than either of the operand, which deviates from the desired behaviour of integer arithmetic.  In an ideal world, an interface for integer arithmetic shall be defined from the expected behaviour, rather than for different integer types of different bit counts, such as \texttt{long} or \texttt{int}

\item The testing of interfaces is generally based on testing cases \cite{unit test}, while most software specification are rule-based. For example, the override of \texttt{Object.equals(Object)} in C\# or Java requires \cite{java} \cite{csharp} are:
\begin{itemize}
\item \emph{The self rule for equality}: an object shall equal to itself.
\item \emph{The reciprocal rule for equality}: The equality testing of object A to B shall equal the equality testing of B to A.
\item \emph{The transfer rule for equality}: If object A equals object B, and B equals object C, then A shall equals C.
\item The other rules.
\end{itemize}
The disparage between the testing framework and the usual requirement framework leads to insufficient tests because it is usually infeasible to provide testing cases to cover all possible scenarios of the rules, especially for procedural programming languages, which is composed of many state machines.  

\end{enumerate}

To address the above deficiency of interface and interface-based programming, this paper introduces a new programming paradigm based on a new behaviour-specification type called a spec.

\section{A Proposed New Way To Specify Functionality}

\subsection{Interface, Trait and Spec}

A \emph{trait} \cite{trait} contains implementations in addition to the what-to-call contract of an interface.  In this respect, an abstract class can also be viewed as a trait.  The implementation in trait may contain more behaviour specification than the function signature of addition, but it may still be insufficient to specify what addition is.

A \emph{spec} is defined here as a trait with specified test cases and test rules.  For example, a spec for addition may be built on the trait for addition, and further requires that:
\begin{itemize}
\item  The result of adding a first operand to a second operand equals the result of adding the second operand to the first operand.

\item  The result of adding two positive operands is larger than any of the operands.

\item  If one operand is a positive integer, the result of the addition equals incrementing the other operand by the integer count.
\end{itemize}

An object implements a spec by 1) implementing all the what-to-call contracts, and 2) passing all test cases and test rules. Because the rule-based behaviour requirement is already coded in the spec, they are tested automatically using call-back techniques, so that a Human developer only need to create spec instances for the test--even this step can be automated when spec instances are collected during a trial run of the spec itself.  Thus, spec also introduces testing rule as its main testing methodology in place of the traditional testing methodology which is based on testing cases.

\subsection{Spec Examples}

A fictional language called \emph{Expresso} is used to describe the new programming paradigm, which uses C\# styled attributes to add properties to each function call. In grammar, a spec looks like a class in an object-oriented language, e.g., each spec instance has a \texttt{this} member to refer to itself, the spec may have public, protected and private access levels for its methods and fields, and a child spec can inherit from a parent spec.  A spec can contain \emph{abstract method} which is declared but not defined method.  A \texttt{ISpec} spec serves as the base spec for all the other specs in Expresso.  \texttt{ISpec} declares an abstract \texttt{IsEqual()} method, as well as two testing rules for the method, which are shown in Figure \ref{ISpec.IsEqual()}.  Each \texttt{[rule]} attribute precedes a testing rule callback method: \texttt{Self()} binds to \texttt{this} and \texttt{Reciprocal()} binds to \texttt{IsEqual(ISpec)}.  During testing, \texttt{Self()} is called for each instance of \texttt{ISpec}, and \texttt{Reciprocal(ISpec)} is called whenever \texttt{IsEqual(ISpec)} is called, to check for the self rule and the reciprocal rule for equality, respectively.  The testing fails if any rule method returns false.  

\begin{figure*}
\begin{verbatim}
[abstract] public Bool IsEqual(const ISpec) const;
[rule(this)] Bool Self() { return this.IsEqual(this); }
[rule(IsEqual(const ISpec))] Bool Reciprocal(const ISpec other) {
    return (this.IsEqual(other)).IsEqual(other.IsEqual(this)); }
\end{verbatim}
\caption{The Declaration for ISpec.IsEqual().}
\label{ISpec.IsEqual()}
\end{figure*}

\texttt{ISpec} spec has another abstract method \texttt{ToString()} to help identify an instance of ISpec, or to represent an instance of ISpec as a character stream, which is shown by Figure \ref{ISpec.ToString()}.  The testing rule \texttt{ToStringReciprocal()} also binds to \texttt{IsEqual()} so that it needs to have different name from the testing rule \texttt{Reciprocal()} in Figure \ref{ISpec.IsEqual()}, because a test rule is not considered as part of \texttt{ISpec}, instead it belongs to the method under test. In Figure \ref{ISpec.ToString()}, the testing rule suggests that \texttt{ToString()} represent an instance of ISpec as a character stream, rather than merely identifying an instance of ISpec.

\begin{figure*}
\begin{verbatim}
[abstract] public String ToString() const;
[rule(IsEqual(const ISpec))] Bool ToStringReciprocal(ISpec other) {
    return (this.IsEqual(other)).IsEqual((this.ToString()).IsEqual(other.ToString())); }
\end{verbatim}
\caption{The Declaration for ISpec.ToString().}
\label{ISpec.ToString()}
\end{figure*}

Before a spec can generate test cases, it needs the ability to be instantiated.  Unlike class constructors, the \emph{spec constructors} are public static abstract methods, as entry-points for the abstract factory \cite{factory} which creates the spec.  A spec constructor which accepts a string as its only parameter is named as a \emph{literal constructor} in this paper, which allows the syntax \texttt{<Spec> <instance>=<literal>} in the code to instantiate the spec.  The literal constructor of \texttt{ISpec} is declared in Figure \ref{ISpec.Create(String)}. When the literal constructor is called during realization test, the attribute \texttt{[case(this)]} generates a test case for all the \texttt{[rule(this)]} test rules, which in turn triggers the \texttt{[rule(IsEqual(const ISpec))]} test rule in a chain reaction.  The attribute \texttt{[rule(Create(const String), Create(const String))]} binds \texttt{Mutual()} test to each mutual pair of multiple \texttt{Create(const String)} calls. 

\begin{figure*}
\begin{verbatim}
[case(this)] [abstract] static public Create(const String);
[rule(Create(const String), Create(const String))] Bool Mutual(const String a, const String b) {
    if (a.IsEqual(b)) return Create(a).IsEqual(Create(b);
    return true;  }
\end{verbatim}
\caption{The Declaration for the literal constructor for ISpec.}
\label{ISpec.Create(String)}
\end{figure*}

\texttt{ISpec} also declares a destructor, a default constructor and a copy constructor with test cases and test rules respectively, as shown in Figure \ref{ISpec constructors and destructor}.

\begin{figure*}
\begin{verbatim}
[case(this)] [abstract] static public ISpec Create();
[case(this)] [abstract] static public ISpec Create(const ISpec);
[rule(ISpec(const ISpec))] Bool Self(const ISpec other) { 
    return ISpec(other).IsEqual(other); }
[rule(ISpec(const ISpec), ISpec(const ISpec))] Bool Mutual(const ISpec a, const ISpec b) {
    if(a.IsEqual(b)) return ISpec(a).IsEqual(ISpec(b));
    if(ISpec(a).IsEqual(ISpec(b))) return false;  // a != b
    return true; }
[abstract] public static Delete(ISpec);
\end{verbatim}
\caption{The Declaration for the default constructor, the copy constructor and the destructor for ISpec.}
\label{ISpec constructors and destructor}
\end{figure*}

\texttt{Bool} implements Boolean arithmetic, and it is a spec derived from ISpec. \texttt{Bool} has two constants defined in Figure \ref{Bool instances}.  The \texttt{Exclusive()} testing rule mandates a \texttt{Bool} instance can be either \texttt{true} or \texttt{false}, but nothing else, which makes the implementation of the abstract methods very simple, such as \texttt{ToString()} shown in Figure \ref{Bool.ToString()}. Each \texttt{[case]} in Figure \ref{Bool.ToString()} attribute precedes a test case, which is implemented by an anonymous method. The \texttt{[override]} attribute in Figure \ref{Bool.ToString()} suggests that \texttt{Bool.ToString()} overrides \texttt{ISpec.ToString()} and inherits all the tests of \texttt{ISpec.ToString()}.

\begin{figure*}
\begin{verbatim}
[case(this)] public static const Bool true("true");
[case(this)] public static const Bool false("false");
[rule(this)] Bool Exclusive() {
    if((this.IsEqual(true)) return true;
    if((this.IsEqual(false)) return true;
    return false;  }
\end{verbatim}
\caption{The Definitions for Bool Instances.}
\label{Bool instances}
\end{figure*}

\begin{figure*}
\begin{verbatim}
[override] public String ToString() const {
    if (this.IsEqual(true)) return "true";
    return "false"; }
[case] Bool () { return true.ToString().IsEqual("true"); }
[case] Bool () { return false.ToString().IsEqual("false"); }
\end{verbatim}
\caption{The Definition of Bool.ToString().}
\label{Bool.ToString()}
\end{figure*}

In \texttt{Expresso}, \emph{method signature can also be overrided}.  In Figure \ref{Bool.IsEqual()}, the attribute \texttt{[override(ISpec.IsEqual(const ISpec))]} states that the argument of \texttt{Bool.IsEqual()} is changed from \texttt{ISpec} to \texttt{Bool}.  Figure \ref{Bool.IsEqual()} also contains a new abstract static method \texttt{Bool(ISpec)} which converts an \texttt{ISpec} instance into an \texttt{Bool} instance, whose \texttt{[auto]} attribute suggests that the compiler may call it whenever such conversion is required, such as when \texttt{Bool.IsEqual(const ISpec)} is called.  The \texttt{domain} attribute will be discussed later.

\begin{figure*}
\begin{verbatim}
[abstract] [auto] static public Bool Bool(const ISpec) const;
[abstract] [override(ISpec.IsEqual(const ISpec))] [domain(object)] public Bool IsEqual(const Bool other);
\end{verbatim}
\caption{The Definition of Bool.IsEqual(const Bool).}
\label{Bool.IsEqual()}
\end{figure*}

In \texttt{Expresso}, \emph{static method can also be overrided}, whose mechanism will be discussed later.
The constructors and the destructor of \texttt{Bool} are shown in Figure \ref{Bool constructor}, in which;
\begin{itemize}
\item  the \texttt{[override(Create(String))]} attribute states that \\ \texttt{Bool.Create(String)} overrides \texttt{ISpec.Create(String)};

\item  the \texttt{[override(Create(const ISpec))]} attribute states that the signature of the copy constructor has been changed from \texttt{ISpec.Create(const ISpec)} to \texttt{Bool.Create(const Bool)}; and

\item  the \texttt{[override(Delete(ISpec))]} states that the method signature has be changed from \texttt{ISpec.Delete(ISpec)} to \texttt{Bool.Delete(Bool)}.
\end{itemize}

\begin{figure*}
\begin{verbatim}
[override(Create(const String))] [domain(object)] static public Bool Create(const String);
[override(Create(const ISpec))] static public Bool Create(const Bool other) {
    if (other.IsEqual(true)) return Bool.Create(true.ToString());
    return Bool.Create(false.ToString()); }
[override(Create())] static public Create() { return Create(false.ToString()); }
[override(Delete(ISpec))] [domain(object)] static public Delete(Bool);
\end{verbatim}
\caption{The definition of the literal constructor, the copy constructor, the default constructor and the destructor for \texttt{Bool}.}
\label{Bool constructor}
\end{figure*}

Although \texttt{Bool} is derived from \texttt{ISpec}, \texttt{ISpec.IsEqual(ISpec)} returns \texttt{Bool} already, because the \emph{method signature} in the new programming paradigm excludes the method return type.  Such seemly peculiarity also exists in other languages, such as between \texttt{Bool} and \texttt{Object} types in C\# \cite{csharp}.  The real problem here is the existence of the \texttt{if} flow control before \texttt{Bool} is defined, whose mechanism will be discussed later.

\subsection{Spec Hierarchy}

The conventional class hierarchy \cite{runtime} has little constraints on behaviour specification.  For an example, when an Apple class is to be derived from a Fruit class, there is little mechanism to guarantee that the Apple class behaves like an apple but not an orange.  Such lax in definition may bring about unexpected behaviours at runtime.  Another problem is that there is no compatibility test when combing parent classes to create a child class.  For an example, methods of the same signature but different intentions and functionalities may exist in multiple parent classes, which result in the implementation-dependent inheritance behaviour \cite{c++} or the diamond inheritance problem \cite{diamond inheritance} of the conventional class hierarchy.  

The spec hierarchy in the behaviour domain is much stricter than the class hierarchy:
\begin{itemize}
\item As a strong-typed system, the method definition of each spec instance can not change at runtime, and each spec has a corresponding \emph{type}, which can be obtained by the \emph{typeof} operator.

\item  A spec may contain either instance or static methods, whose access level can be either public, protected or private to the spec.  

\item  The \emph{specific name} for each method is $<$spec name$>$.$<$method name$>$, in which $<$spec name$>$ is the name of the spec which defines the method.  If a method signature along can not uniquely identify a method in a context, the method has to be identified by its specific name.     

\item  A spec uses test cases and test rules to specify the desired behaviours for its methods.  

\item  A child spec can be derived from one parent spec or multiple parent specs. 

\item  When a child inherits a method from a parent spec, it also inherits all the test cases and test rules for the method. A child spec can use the \texttt{violate} attribute to bypass a specific testing rules of its parents, while a parent spec can prevent such bypass by using the \texttt{must} attribute.  For example, the two rules in Figure \ref{ISpec.IsEqual()} seem to be good candidates for having \texttt{must} attribute, while the rule in Figure \ref{ISpec.ToString()} may be potentially violated when \texttt{ToString()} is merely used to identify an spec instance.  A child spec can add new tests, however, the parent spec of which the method is inherited from needs not pass these new tests.

\item If two specs conflict with each other in their tests, they can not be both parent specs for a child spec, unless the child spec uses the \texttt{violate} attribute to resolve such conflicts.

\item  A child spec can restrict the access level for an inherited method, such as from protected to private. When a method is inherited from multiple parent specs, the access level for the method is the most restrictive among all the parent specs.

\item  Using \texttt{override} attribute, a child spec can override any method which it inherit, by providing a new definition.  The \texttt{override} attribute also allows the argument types of the method be overrided to a child spec of each of the original argument spec, provided that such conversion is defined, and further labelled as \texttt{auto}.

\item  A parent can label its method \texttt{final}, preventing the method from being overrided.

\item  A child spec can inherit multiple method definitions of the same method signature from different parent specs.  If all the inherited methods have the same specific name, they are from the same spec source and aggregated as one method definition for the child spec, with tests from all the parents combined; otherwise, the method becomes abstract for the child spec, which the child spec may override. Thus diamond inheritance problem \cite{diamond inheritance} is avoided. From business perspective, a method is like a clause in a contract set \cite{business law} so that each method shall have only one definition everywhere.

\item  A spec may contain either instance or static private fields, which are only exposed using method calls, such as getters and setters.  
\end{itemize}

A \emph{method table} maps each method signature to a method definition, or nothing when the method is abstract.  The method definition also contains other characteristics for the method, such as its access level and its return spec. 
\begin{itemize}
\item  Each spec has a corresponding method table called a \emph{native table} which maps all its methods.  The native table is built following the inheritance generations, so that the definition provided by a child spec always overriding the definition provided by its parent spec. 

\item  When a child spec is cast to a parent spec, a method table called a \emph{cast table} is constructed by mapping each method signature in the native spec table of the parent spec to the corresponding method definition in the native spec table of the child spec.  For example, Figure \ref{(ISpec)Bool} contains the cast table when \texttt{Bool} is cast to \texttt{ISpec}. 

\item  \texttt{ISpec} has a static method \texttt{this} which always returns the native table of the spec, and which is always the starting method table for any casting.  
\end{itemize}

\begin{figure*}
\begin{tabular}{l l}
Signature & Definition \\
\hline
\texttt{ToString() const} & \texttt{public String Bool.ToString() const} \\ 
\texttt{IsEqual(const ISpec) const} & \texttt{public Bool Bool.IsEqual(const Bool) const} \\
\texttt{static Create(const String)} & \texttt{[domain(object)] static public Bool Bool.Create(const String)} \\
\texttt{static Create(const ISpec)} & \texttt{Bool Bool.Bool(const Bool)} \\
\texttt{static Create()} & \texttt{static public Bool Bool.Create()} \\
\texttt{static Delete(ISpec)} & \texttt{[domain(object)] static public Void Bool.Delete(Bool)} \\
\end{tabular}
\caption{The cast table for \texttt{(ISpec)Bool}.}
\label{(ISpec)Bool}
\end{figure*}

Each spec is characterized by a \emph{parent table} which maps each of its parent name to a corresponding cast table, in addition to mapping \texttt{this} to the native table.  When a spec is cast to anther spec, the corresponding cast table is fetched from the parent table, or an exception may be raise if such casting is not defined in the parent tabble. 

A \emph{spec table} maps each spec name to a corresponding parent table, which is used to create each spec.

\subsection{Resolution of Method Overriding}

\begin{figure*}
\begin{verbatim}
class B { public virtual String F() { return "B.F()"; } }
class C : B {}
class D : B { public override String F() { return "D.F()"; } }
class E : D {}
\end{verbatim}
\caption{Typical C\# inheritance codes which contains method overriding.}
\label{dynamic overriding}
\end{figure*}

\begin{figure*}
\begin{verbatim}
String __F__( Type type, B* pThis ) {
    if (type == typeof(E)) { ((E*) pThis)->F(); }
    else if (type == typeof(D)) { ((D*) pThis)->F(); }
    else if (type == typeof(C)) { ((C*) pThis)->F(); }
    else { pThis->F() }
}
\end{verbatim}
\caption{C-table implementation of method overriding.}
\label{c-table overriding}
\end{figure*}

Method overriding is traditionally implemented using v-tables, which looks for the method definition by the virtual method signature for each class \cite{v-table}. For example, to implement method override in Figure \ref{dynamic overriding} using v-table:
\begin{enumerate}
\item  Each instance contains a hidden pointer to a v-table, which is specific for each class, mapping virtual method signature to definition, such as mapping the virtual method \texttt{F()} to \texttt{B.F()} for class \texttt{B}.

\item  When a child class is derived from a parent class, the child class first makes a copy of the v-table of the parent class, then adding new virtual method definitions or replacing existing method definitions, such as mapping \texttt{F()} to \texttt{D.F()} instead of \texttt{B.F()} in class \texttt{D} in Figure \ref{dynamic overriding}.

\item  At run time, when a virtual function is called on an object, the v-table of the actual class is searched for the method definition before it is called.  For example, in Figure \ref{dynamic overriding}, when \texttt{F()} is called, instances for class \texttt{B} and class \texttt{C} will use \texttt{B.F()}, while instances for class \texttt{D} and class \texttt{E} will use \texttt{D.F()}, regardless if instances for class \texttt{D} and class \texttt{E} appear as instances for class \texttt{B} and class \texttt{C}.
\end{enumerate}

Because a v-table is associated with each instance of a class, static methods can not be virtual. A run-time cost of method definition look-up and indirect call is incurred for each virtual method call, which may disrupt instruction pipeline for a modern CPU. Virtual method call can only be supported by programming languages which supports v-table, so that complicated mechanism, such as COBRA or COM, is needed to support method overriding functionality in legacy programming languages. 

In contrast, the new programming paradigm implements method overriding using a new technique called c-table, which looks for the method definition by class for each virtual method signature. For example, to implement method override in Figure \ref{dynamic overriding} using c-table:
\begin{enumerate}
\item  Insert a c-table for each virtual method, which is a table mapping native spec name to definition, such as mapping class \texttt{B} to \texttt{B.F()} for the virtual method \texttt{F()}.

\item  Populate the c-table with the child class name to the existing method definition, such as mapping from class \texttt{C} to \texttt{B.F()}.

\item  Replace the definition in c-table when a child class override the virtual method definition, such as mapping \texttt{D} to \texttt{D.F()} instead of \texttt{B.F()} in the c-table for \texttt{F()}.  

\item  When calling a method from a spec at run-time, use the type of the native spec to find the method definition, such as reaching \texttt{B.F()} from class \texttt{B} or class \texttt{C}, while \texttt{D.F()} from class \texttt{D} or class \texttt{E}.
\end{enumerate}

Implementing method overriding using either v-table or the c-table only differs in dimensions of method signature v.s. class name.  Similar techniques to optimize v-table from a lookup table into a index table \cite{v-table} can also be used to optimize the c-table. Thus, the runtime speed of resolving method overriding using either technique is comparable.  Moreover, because each c-table collects all the overriding definitions of a virtual method together, it may allows better compile-time optimization.  For example, the c-table lookup code inside the virtual method definition can be replaced with a branch control, with all the overriding definitions possibly inlined in the method, so that the virtual method call can be resolved without pointer dereference, and with the effective help of branch prediction. Figure \ref{c-table overriding} shows such \texttt{C} implementation of method overriding for method \texttt{F()} in Figure \ref{dynamic overriding}, in which a hidden stand-along method \texttt{\_\_F\_\_()} is postulated to be generated by a compiler.  Because each instance no longer carries a v-table pointer, it separates data operation clearly from data definition, which is in line with the new programming paradigm.  For example, in Figure \ref{c-table overriding}, each instance is represented by \texttt{pThis}, which refers to pure data of the instance, such as a structure in \texttt{C}.  Figure \ref{c-table overriding} also shows that the new programming paradigm can be implemented by a procedure language which is not even object oriented, e.g. \texttt{C} or \texttt{ForTran}.  As a main objective of the new programming paradigm, the method resolution using c-table also allows static method to be overrided, which may be used to move reflection methods to be implemented all at compile-time.

\section{A Proposed Separated Programming Domains}

Even though spec introduces a better way to specify functionality, it may still be desirable to adapt the spec hierarchy to exiting programming languages and libraries.  To achieve this goal, a new programming paradigm is proposed to limit spec hierarchy to describe object interactions only. 

\subsection{Domains of Operations}

The operations by any computer program can be divided into three domains:

\begin{itemize}
\item  \emph{Object domain}:  This domain focuses on the creation, destruction, cloning, passing, transmitting and persisting of individual objects, but not on how objects interact with each.    

\item  \emph{Behaviour domain}:  This domain focuses on how objects interact with each, and what are the expected behaviours for the set of exposed method of each object.   Interfaces and traits rather than type have been preferred method to specify behaviours of an object \cite{pragmatics}.  Spec should be a better specification.  The question is if any behaviour can be specified by implementation-neutral specs alone.  When arithmetic operations are performed, the actual value of an arithmetic operand is never used directly but only indirectly through the comparison and the arithmetic relations.  This seems to be a general rule for programs in behaviour domain.  So it may be quite possible to specify behaviour using implementation-neutral specs only.  

\item  \emph{Realization domain}:  This domain focuses on how a behaviour specification in the behaviour domain binds to a concrete object in the object domain.  Such binding is hard-coded when a type which implements an interface is instantiated directly to provide the services of the interface in a traditional programming language.  A preferred behaviour realization is to use the abstract factory pattern \cite{factory}, which provides an implementation to an interface as the interface only.  The new programming paradigm uses specs instead interfaces, and uses the abstract factory pattern \emph{exclusively} to provide the spec implementation.  In this way the codes to manage objects are clearly separated from the rest of the codes, with the former in the object domain and the latter in the behaviour domain.  The realization domain connects the two very different domains of operations, and embeds each entity with dual roles.  For example, a spec may appear in a method signature in the behaviour domain, while a heap pointer of the underlying type may actually be copied into the method in the object domain.
\end{itemize}

The design philosophy of the new programming paradigm is to be as strict as possible in the behaviour domain, while as flexible as possible in the object domain, relying on spec testing in the behaviour domain to validate implementation in the object domain.  In the new programming paradigm, codes in the behaviour domain are strong typed and statically compiled, while codes in the object domain can be dynamically linked.  Such flexibility in the object domain allows runtime creations and changes of the underlying data type definitions, while such strictness in the behaviour domain guarantees the correctness and consistency of the software behaviours.

\subsection{Domain Examples}

In \texttt{Expresso}, \texttt{[domain]} attributes labels a method to be implemented in the object domain.  Because \texttt{Bool}'s constructors create instances of \texttt{Bool}, some of them need to be defined in the object domain, e.g., storing a \texttt{Bool} instance as either an integer (as in C/C++), or a byte or even a bit should be irrelevant on how a \texttt{Bool} instance should behave.  For the same reason, the destructor of \texttt{Bool} also needs to be defined in the object domain.  For example, in Figure \ref{Bool constructor}, the \texttt{[domain(object)]} attribute states that the definition of \texttt{Bool.Create(String)} exists in the object domain.  

The implementation of \texttt{Bool.IsEqual()} also need to be defined in the object domain because it depends on how a \texttt{Bool} is actually stored in the object domain.  \texttt{Bool.IsEqual()} is the base for comparison in general, such as in Figure \ref{ISpec.IsEqual()} and \ref{ISpec.ToString()}.  It is also the base for logic flow constructs, which also need to be defined in the object domain, such as the \texttt{if} clause in Figure \ref{Bool.ToString()}.  

Having only one constructors, one destructor and one method defined in the object domain, \texttt{Bool} represents a behaviour-intensive spec.  In contrast, as the other spec referred by \texttt{ISpec}, \texttt{String} represents an implementation-intensive spec, because different languages even different implementations of the same languages can have very different implementation model for their \texttt{String} type definitions.  It is possible that the new programming paradigm supports a variety of string implementations simultaneously, each of which binds to the \texttt{String} spec by a different parameter set in the factory, such as selecting a string implementation optimized for short size with constant content, or a string implementation optimized for long size with variable content, while both of which have the same behaviour.  So the \texttt{String} spec may contain mostly test rules and test cases, while leaving most implementation to the object domain. 

In theory, \texttt{String.IsEqual()} can be defined in the behaviour domain; however, such definition may be much less efficient than an implementation in the object domain by taking advantage of how a \texttt{String} is actually stored in the object domain, so that it is labelled as \texttt{[domain(dual)]}, with an initial definition provided in the behaviour domain, and to be overrided in the object domain if desired.

\subsection{Compilation and Linking Process}

One of the reasons to separate out behaviour domain is to unify the programming space, which has been fractured by different programming languages, or even different flavours of the same programming languages. The new programming paradigm realizes that different programming languages have different strength and usage histories in managing objects and their associated behaviours, and the behaviour domain should be a common platform for these different programming languages to interact with each other.  Thus, the new programming paradigm will not try to manage object and resource, or rewriting existing behaviour specification of existing programming languages; rather, it will clarify existing behaviour specification using testing rules and testing cases once these behaviour specification has been repackaged to a unified format using \texttt{spec}.  To achieve this goal, the new programming paradigm compilation does not generate executable directly but implores other compilers called \emph{target compiler} to do so.  A \emph{concrete spec} is a spec which is preceded with the \texttt{[concrete]} attribute.  Only concrete specs are instantiated in the object domain.  The \texttt{[concrete]} attribute also specifies the targeted compiler of the compilation process of the new programming paradigm.  For example, the \texttt{Bool} spec can be tagged as:

\begin{verbatim}
[concrete(C)] spec Bool : ISpec {...}
\end{verbatim}

The new programming paradigm is statically typed and statically compiled in the behaviour domain, with most workload done in the compilation process, so that its behaviour is very predictable at runtime.  The new programming paradigm compilation process is the following:
\begin{enumerate}
\item  Generates a native table and a parent table for each spec, as well as a spec table for all specs.  

\item  Populate a c-table for each method.

\item  Generates a traditional interface called a \emph{contractor} for each concrete spec, which is composed of all the methods which are preceded with \texttt{[domain]} attributes.  For example, assuming that \texttt{C} is the targeted language, Figure \ref{Bool contractor} is the contractor for \texttt{Bool}, in which both \texttt{Type} and \texttt{Bool} are \texttt{C} \texttt{struct} generated from the corresponding specs in \texttt{Excresso}. 

\begin{figure*}
\begin{verbatim}
Bool Bool_Create(const char const *)
void Bool_Delete(Bool *)
Bool Bool_IsEqual(const Type const *, const Bool const *, const Bool const *)
\end{verbatim}
\caption{The contractor for \texttt{Bool}.}
\label{Bool contractor}
\end{figure*} 

\item  For each spec instance, generate a \emph{cast pointer} pointing to an entry in the parent table of the spec.  The compilation process fails if a spec instance is cast outside the parent table of the spec.  

\item  Apply the specific name for each method reference.  The compilation process fails if a method is referenced but not defined, or if the access control is violated, or if the specific name can not be uniquely determined.

\item  Generate a \emph{test table} for each concrete spec, which maps each method signature to a set of test rules as well as a set of test cases.  The test rules and the test cases from the parent specs are all added to the corresponding sets in the concrete spec.  

\item  Adapt the spec codes to the targeted compiler and compiles the codes for the concrete specs in the behaviour domain using the targeted compiler.  For each concrete spec, the compiled result is called the \emph{executor}, while the compiled result for the corresponding test table is called the \emph{tester}.

\item  Compiles the codes in the object domain against the contractor.  The compiled result which binds to each contractor is called the \emph{implementer} for the corresponding concrete spec.    

\item  Link executor with corresponding implementer through contractor.

\item  Runs through all the test cases in the tester, which trigger all the test rules in a chain reaction.  If any test fails, the compilation process also fails.  
\end{enumerate}

\begin{figure*}
\centering
\includegraphics[height=2.5in, width=5in]{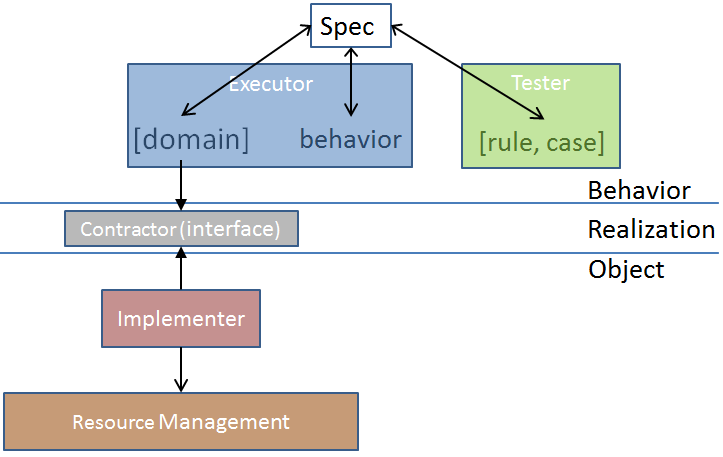}
\captionof{figure}{Domains of operations in the new programming diagram.  }
\label{fig: Domains}
\end{figure*}

Figure \ref{fig: Domains} shows the players of the new programming paradigm when the programming space is separated into a behaviour domain, an object domain, and a realization domain in between:
\begin{itemize}
\item   The tester in the behaviour domain contains those codes labelled by \texttt{[rule]} and \texttt{[case]} attributes, which calls the executer in the behaviour domain.

\item  Though the contractor in the realization domain, the executer in the behaviour domain calls the implementer in the object domain, with the calling scope defined by the \texttt{[domain]} attributes in the behaviour domain. 

\item  The implementer interacts with operating system directly to manage resources.  The work-flow logic such as \texttt{if} logic is also provided by corresponding implementer for the programming language in the behaviour domain.  

\end{itemize}

A contractor can be implemented by multiple implementers in the object domain, one of which can be chosen at runtime through either a parametrized spec constructor or a configuration file in the object domain.  For an example, an a parametrized \texttt{String} constructor may choose among different string models for the intended usage.  Such dynamic binding scheme also makes the debugging and enhancement of code easier.  For example, a change of implementation such as addition of new members to a data object is transparent to the behaviour domain codes.

Thus, the new programming paradigm is aimed to be a general-purposed, statically compiled \cite{pragmatics} strong-typed \cite{pragmatics}, dynamically linked \cite{pragmatics}, self-tested \cite{unit test}, multi-platform, multi-target and source-to-source \cite{source-to-source} programming language. To achieve this goal, a common protocol for data exchange between different programming languages seems required, which is described in the next section.

\section{Type Expressiveness}

One difficulty of the new programming paradigm is how to let software written in different languages running on different platforms calling each other.  Instead of the traditional protocol such as COM or COBRA, this paper propose a new method called type expressiveness, in which each type instance reveals how to construct it using a common protocol.  The type expressiveness allow parameter passing and result passing, thus enables cross-language and cross-platform call possible, once mapping of primary types between two computer languages are provided.

\subsection{Definition}

\emph{Type expressiveness} is defined as using public getter and related attributes to reveal each type's information in a self-contained way, so that sufficient information to clone the instance can be obtained by querying public getters in a systematic manner.  If a type is \emph{fully expressive}, the clone will be identical to the source.  Although type expressiveness is demonstrated using C\# in this paper, same programming principles can be applied to other programming languages.  

A few types are predefined to be \emph{primitively expressive}, whose values can be converted between the corresponding string values:
\begin{enumerate}
\item  A C\# string is primitively expressive.  

\item  Any C\# primitive type \cite{csharp} is primitively expressive.  Each primitive type implements 1) an instance \texttt{ToString()} method to convert its value to a string, and 2) a static \texttt{Parse()} method to construct an instance of the type from a value-coded string.

\item  Any C\# value type \cite{csharp} is assumed to be primitively expressive if the type 1) has override \texttt{ToString()}, and 2) has a static \texttt{Parse()} method that accepts a string parameter and return a instance of the value type.  This category includes all the C\# enum types.  It also includes other commonly used C\# value types such as \texttt{DateTime}.  
\end{enumerate}

The following are some common good practices regarding to data encapsulation in C\# \cite{effective csharp}
\begin{itemize}
\item  Field should always be private or protected in a class unless the field is a public constant.

\item  Field should usually be accessed through getter and setter.  Public setter should only exist if it is absolutely necessary to do so.

\item  Field should be read-only whenever possible.  A read-only field is assigned either in constructor or when the field is defined and initiated, and its value is no longer allowed to change after the initial assignment.  

\item  To better handle exception, sometimes the construction of an object is divided into two steps: 1) calling a constructor, and 2) calling one or several initiation methods immediate after calling the constructor.
\end{itemize}

Following the above common good practice, a type is expressive if its public getters are sufficient to reveal how it is constructed, namely:
\begin{enumerate}
\item  It has either a public default constructor, or an expressive constructor.  An \emph{expressive constructor} is a public constructor with each parameter matching a public getter.  The matching requires that 1) the type of the getter equals the type of the parameter, and 2) the names of the getter and the parameter matches according to predefined matching rules.  The default matching rules require that the two names have to be identical after masking off camel case, type decorator, and collection decorator.

\item  It may have some expressive initiation methods.  An \emph{expressive initiation method} is a public method which 1) receives no parameter, and 2) is tagged by an \texttt{[InitSeq(int)]} attribute.  The parameter in the \texttt{[InitSeq(int)]} attribute defines the calling sequence if there are multiple initiation methods.  

\item  It may have some instance expressive setters or static expressive setters.  An \emph{expressive setter} is a public setter which 1) has a corresponding public getter, and 2) is not tagged by the \texttt{[NonExpressive]} attribute.

\item  It may have some instance expressive getters or static expressive getters.  An \emph{expressive getter} is a public getter which 1) has no corresponding public setter, 2)  is not tagged by a \texttt{[NonExpressive]} attribute, and 3) is a collection implementing the \texttt{ICollection} interface.  An expressive getter is used to serialize the elements of the collection.  
\end{enumerate}

\begin{figure*}
\begin{verbatim}
interface MyIntface {
    int MyValue { get; set; }
    bool ConnectToDB();
}
public class MyExpressiveType : MyIntface {
     // an internal field that can construct itself needs not be expressive 
     private static readonly ILog myLog = LogManager.GetLogger("My Log");

     // an expressive setter
     public double MyValue { get; set; }

     // an expressive getter
     public List<double> MyCollection { get { return myCollection; } } 
     private readonly List<double> myCollection = new List<double>();

     // a non-expressive getter
     [NonExpressive]  // a marker for non-expressiveness
     public double[] MyArray { get {
        double[] clone = new double[MyCollection.Count];  myCollection.CopyTo(clone, 0);  return clone; } }

     // an expressive constructor
     public MyExpressiveType(double myReadonly) {  // match to getter not field
         this.myConnectionString = Encryptor.Decode(myReadonly); }
     public double MyReadonly { get {  // use camel case
         return Encryptor.Encode(myConnectionString); } }
     private readonly string myConnectionString;

     // two expressive initiation methods 
     [Init(1)]  // return type suggests that first call can fail
     public bool ConnectToDB() { 
         myConnection = DB.Connect(myConnectionString, MyValue);
         return myCollection != null; } 

     [Init(2)]  // second call depends on previous successful calls
     public void SetDBLogger() { myConnection.SetLogger(myLog); } 

     private DBConnection myConnection;
}

\end{verbatim}
\caption{An Example Expressive Type}
\label{expressive type}
\end{figure*}

Figure \ref{expressive type} displays a simple example of a fully expressive type definition, which also shows that expressiveness of a type can be used to document the behaviour of the type from the perspective of the minimal information to recreate an instance of the type, e.g., the \texttt{[NonExpressive]} attribute suggests that an array getter will not expose its underlying collection, and the two \texttt{[InitSeq]} attributes suggest the dependency relation between the two initiation methods.

The major improvement of the serialization using type expressiveness over the default .NET XML serialization functionality is data encapsulation.  The default .NET XML serialization functionality 1) always constructs an object using the default constructor, and 2) ignores all the getters without the corresponding setters, so that the field to be serialized can not be read-only.  Instead, using type expressiveness, type expressiveness can be constructed in the expressive constructor, so that it can be read-only, and without corresponding setter.

\subsection{Default Naming Convention for Constructor Parameters}

One of the major efforts to make a type expressive is to match the name of each parameter in an expressive public constructor with a name of a public getter.  Such name matching already exists in common practices, such as the camel naming convention \cite{effective csharp} between property and field, or the usual identical name between a constructor parameter and a read-only field [19].  These common practices are formalized as the default naming convention for the expressive constructor parameters, in which a parameter name is allowed to have the following optional difference with the corresponding getter name:
\begin{itemize}
\item Camel case:  The first letter of a parameter name can be the small letter of the first capital letter of a getter name.

\item Type decorator:  An extra character can precede the name of a parameter to decorate the type of the parameter, such as \texttt{"i"} for signed integer, \texttt{"u"} for unsigned integer, \texttt{"d"} for double, \texttt{"w"} for string, \texttt{"b"} for Boolean.

\item  Array/Collection decorator:  An extra \texttt{"s"} can follow the name of a parameter to indicate that the parameter is an array or a collection.
\end{itemize}
After decorators are removed, and camel case is corrected, the parameter name matches a getter name if they are case-sensitive equal.

\subsection{Serialization of Expressive Types}
\label{stream type}

A 8-bit char stream seems to be the only data format that can be ubiquitously understood among the different languages and/or among on different operating systems, e.g., regardless of endianness \cite{pragmatics}.  With a proper ending protocol, a string can be translated among the different languages and/or among different operating systems.  Thus, instance of string can be regarded as the basic transferable elements in a stream between a sender and a receiver in general cases of serialization \cite{serialization}.  All strong typed languages tend to have similar set of primitive types, and the same set of string representation of these primitive types \cite{pragmatics}.  Because all other types are made from string and primitive types \cite{pragmatics}, instances of any type can be coded into the stream.  In additional to instance data, an \emph{expressive stream} also contains type information and name for each type instance in the stream.  The expressive stream may contain the following types of stream items:
\begin{itemize}
\item  \texttt{VString} packs an instance name with a string.  

\item  \texttt{Prim} packs an instance name with a value of primitively expressive type other than a string.  

\item  \texttt{Value} packs an instance name for a stack object in C\#. 

\item  \texttt{Refer} packs an instance name with a hash code for a heap object.  A zero hash code means that the instance is a null heap object.  

\item  \texttt{TypeInfo} packs a type name.  

\item  \texttt{IntfInfo} packs a type name with an interface name. 
 
\end{itemize}
It is postulated that the expressive stream may also contains the \texttt{Code} type of stream items which embed \texttt{Expresso} codes for initiation methods in the same way as a HTML document embeds JavaScript code \cite{java script}.  

The order of stream items for an instance of non-primitively expressive type is the following in an expressive stream.  
\begin{enumerate}
\item  A \texttt{TypeInfo} item for the type.

\item  A \texttt{IntfInfo} item for each implemented interface of the type, which exist at most once for each type.

\item  A stream item for each static expressive setter, which exists at most once for each type.

\item  Stream items for each collection item of each static expressive getter, which exist at most once for each type.

\item  A \texttt{Value} item for each stack object or a \texttt{Refer} item for each heap object.  

\item  A stream item for each instance expressive setters.

\item  A stream item for each expressive constructor parameter.

\item  Stream items for each collection item of each instance expressive getter.

\end{enumerate}
Such scheme applies regressively to instances of constructor parameters, expressive setters and collection items of expressive getters, until all objects are reduced to instances of primitively expressive types.

\subsection{Type Binding in Expressive Stream}

The above step 1 to 4 in the serialization process applies \emph{type binding} to each instance of the type which is not primitively expressive type.  To apply type binding to an instance of a primitively expressive type, a \texttt{TypeInfo} item precedes the instance.  In the complete format, an expressive stream applies type binding to all type instances which it contains.

\subsection{Name Binding in Expressive Stream}

In \emph{name binding}, the name for each stream item is derived from either type name or instance name using the following conventions:
\begin{itemize}

\item  \texttt{<type name>:<interface name>} identifies the name of an interface for the type.

\item  \texttt{<type name>.<setter name>} identifies the name of a static expressive setter for the type.

\item  \texttt{<instance name>.<setter name>} identifies the name of an instance expressive setter for the type instance.

\item  \texttt{<instance name>\$<getter name>} identifies the name of an instance getter which acts as a parameter in an expressive constructor for a type instance.

\item  \texttt{<type name or instance name>*<getter name>} identifies the name of a static or instance expressive getter for a type when the expressive getter is either an array or an \texttt{IList} collection.

\item  \texttt{<type name or instance name>@<getter name>} and \texttt{<type name or instance name>\&<getter name>} identify the name of a static or instance expressive getter for a type when the expressive getter is a \texttt{IDictionary} collection and the stream item is the key and value for the \texttt{IDictionary} collection, respectively.

\end{itemize}
Name binding also applies regressive to each stream item.

\subsection{Deserialization of Expressive Types}

Type and name binding in expressive stream is necessary because the receiving end of the stream may be written by a different software version or even a different language.  Using type and name binding, a complete data set of an instance of a non-primitive expressive type is retrieved and reordered from an expressive stream.  The instance is then reconstructed in the following order:
\begin{enumerate}
\item  Check to see if the system has the serialized type defined.  If so, match the serialized type with the defined type; otherwise, the type needs to be created dynamically.  

\item  If the type's interface set has never been checked for the type, check its interface set to make sure that the type has identical interface set in the stream.

\item  If the type's static expressive setters have never been called for the type, call the static expressive setters.

\item  If the type's static expressive getters have never been populated for the type, populate the static expressive getters.

\item  Call the type's expressive constructor.

\item  Call the type's instance expressive setters.

\item  Call the type's expressive initiation methods in sequence.  If a method returns false, stop the construction, and throw an exception.  Other return types of the initiation method are ignored.  

\item  Call the type's instance expressive setters again.

\item  Populate the type's instance expressive getters.
\end{enumerate}
Such scheme applies regressively in the order exactly opposite to the serialization process.

To better handle different versions of a same type definition, a \texttt{def} method may be introduced in addition to the \texttt{get} and \texttt{set} methods for property, which hold the default value of the property.

\subsection{Reference Integrity}

All objects are cloned after a serialization and de-serialization cycle.  A heap object is access through a reference, which has one-to-one correspondence with hash code for the heap object in C\# and Java.  If there are multiple references to a heap object, the reference relationship is preserved among de-serialized objects.  This is called the \emph{reference integrity} \cite{referential integrity}.  

To achieve reference integrity, all serialized references are saved in a table indexed by the hash codes.  During serialization, if a reference is not in the table, a full heap object follows the \texttt{Stream.Refer} item in the stream; otherwise only \texttt{Stream.Refer} item exists in the stream.  During de-serialization, when a \\ \texttt{Stream.Refer} object is fetched from the stream, the table is populated if it does not contain the reference, and all subsequent \texttt{Stream.Refer} object in the stream gets the reference from the table.

\subsection{Format Difference of Expressive Streams}

An expressive stream is not always in the complete format.  When the sender and the receiver shares the same type definition, the member types of each type is known on both ends of the stream so that the type binding may not be necessary for the members whose types are sealed.  The order of the stream is also known on both ends of the stream so that the name binding for members may also be omitted.  The simplified expressive stream is much more efficient in transferring type instances between the sender and the receiver.  A special case is that the sender is the receiver, which results in cloning.  

The format of each expressive stream is set when the stream is constructed, so that each expressive stream has only one format.

\subsection{Checking for Full Expressiveness}

If after an instance of a type is cloned using expressive stream, and the clone instance equals the original instance, the type is \emph{fully expressive}.

Instead of overriding \texttt{Equals()} for each type, the concept of type expressiveness can also be used to compare the two instances after cloning.  Using reflection, each object can be decomposed into fields of primitively expressive type, and then compared field-by-field.  The comparison is carried out recursively to all the fields and all the base classes unless specified otherwise: 
\begin{itemize}
\item  If a field is tagged with the \texttt{[CompareIgnore]} attribute, it is ignored for comparison.

\item  If a base class is tagged with the \texttt{[CompareBase]} attribute, the comparison will not go deeper than that class. A child class can also use the \texttt{[CompareBase(String)]} attribute to specify the base class to stop the comparison.  
\end{itemize}

\subsection{A Persistence Example of Expressive Types}

\begin{figure*}
\begin{tabular}{l l l l l}
\#? & UnitTest.MyExpressiveType & UnitTest.MyIntface &
System.Double &	System.Int32 \\
\#@ & MyExpressiveType & MyExpressiveType:MyIntface & MyExpressiveType\$MyReadonly & MyExpressiveType.MyValue \\
? & 19342748 & UnitTest.MyExpressiveType  & 12345 & 6789 \\ 
\#? & System.Double \\
\#@ & MyExpressiveType*MyCollection \\
? & 0.123 \\
? & 456.7 \\
? & 890 \\
\end{tabular}
\caption{The Tab-Delimited File Format for an instance of the Example Expressive Type of Figure  \ref{expressive type}.  }
\label{expressive file}
\end{figure*}

Data frequently needs to be in delimited format to be exchanged with other data processing programs, such as the Microsoft Excel and the R Statistics Package.  Thus a delimited format for serialization is quite useful.  An expressive and serializable tab-delimited text file contains the following three special types of lines in addition to normal text:
\begin{enumerate}
\item  \emph{type line}:  ``\#?\textbackslash t\textless  Type1\textgreater \textbackslash t\textless  Type2\textgreater ...\textbackslash n''.  A type line must leads any other formatted text lines.

\item  \emph{name line}:  ``\#@\textbackslash t\textless  Name1\textgreater \textbackslash t\textless  Name2\textgreater ...\textbackslash n''. A name line must follow a type line, with the delimited fields in both lines matched by the column positions.

\item  \emph{data line}:  ``?\textbackslash t\textless  Datum1\textgreater \textbackslash t\textless  Datum2\textgreater ...\textbackslash n''. One or multiple data lines must follow each pair of type and name lines. The type and name of each delimited field in the data line is provided by the corresponding delimited fields in the type line and the name line of the same column position. 
\end{enumerate}
Figure \ref{expressive file} shows the tab-delimited file format for an instance of the example expressive type of Figure \ref{expressive type}

An expressive stream contain all information to generate the tab-delimited text files, except the format information, such as how to break stream into lines and files, and in what sequence to write the stream into files.  Such decisions should be made by the Human user who generates the files. Because the de-serialization process reorders the fully expressive stream according to type and name binding, it can easily handle expressive stream in any persisted formats. 

\subsection*{Prospective}

The separation of object domain and behaviour domain may provide a new dimension of data abstraction.  An expressive type can be generated dynamically from an expressive stream, which can exists as an actual stream, as a file of various formats, the content of a web page, or even as data in a database. The instance of the expressive type is manipulated according to different interfaces of the type in the behaviour domain, each of which provide one slice of the expressive type.  Thus, there may be no need to write code manually to represent each expressive type but to generate the code dynamically from 1) the data existence in the object domain and 2) the data usage in the behaviour domain.  For example, in java spring framework \cite{java spring}, each property in the bean file corresponds to a setter.  From the perspective of type expressiveness, if the type of the property is also provided, the code for the setter becomes redundant, which can be generated dynamically at runtime.  For example, the code in Figure \ref{expressive type} can almost be generated by the data in Figure \ref{expressive file}.  In the programming paradigm, an object in the object domain represents a pure data object, which should be contains no or minimal behaviour codes, or even be determined dynamically by the data in various expressive formats.  Each data file can be validated by various testers before it is put into production so that this dynamic linking approach may be quite safe.

\section{Summary and Discussion}

The main idea of the new programming paradigm seems simple and straight-forward: 1) to use spec instead of interface for behaviour specification; 2) to separate programming space into behaviour domain, object domain and realization domain, and 3) to use the expressive stream to transfer objects among different address spaces.  More details needs to be discussed and laid out.  

The new programming paradigm is just a concept at this moment. It still has a lot of important questions to be answered.  For instance:  
\begin{itemize}
\item  It is not clear how to specify a multi-thread performance requirement using test rules and test cases of the specs.  Threading in different programming languages on different operating systems is an implementation-intensive issue, e.g., it has been reported that Java threading is less efficient than C\# threading on Windows platform due to JVM abstraction of threading on all platform, including those which actually do not support threading \cite{thread: csharp vs java}.  How to effectively abstract threading in behaviour domain remains a challenge.  

\item  It may be necessary to distinguish between compile time test and runtime test. with the former for behaviour test on a build machine and the latter for performance test on a target machine.  

\item  It is not clear how to constraints the outcomes using specs because test rules and test cases of the specs are essentially tools to specify what are the desired behaviours; however they are not good at rejecting the undesired behaviours or side-effects such as all the possible parasite behaviours under the influence of computer viruses  \cite{virus}.  

\item  The static compilation and dynamic linking aspects of the new programming paradigm calls for a security scheme to reliably certify an implementer once the implementer has been tested against the contractor and the tester.  Because the testing can be execution-intensive and time-consuming, runtime testing of the implementer may not be practical.  Also, the test of an implementer can be done on a powerful development machine, while the target machine on which the implementer is deployed can be a machine with very limited resources, such as an embedded system.  Thus, a implementer may need to be digitally certified in one machine before it is shipped to another machine.  How to prevent a illegitimate impersonator to mimic a legitimate implementer remains a challenge.

\item  In contract with the conventional unit tests \cite{unit test}, in which all test cases are activated sequentially, in the new programming paradigm, the test rules and test cases are arranged by methods so that each test rule may be activated multiple times for the same test case.  How to activate the test rules more efficiently at compile time remains yet another challenge.

\item  The biggest challenge is still how to integrate software components in different computer languages running in different environments.  The current implementation of type expressiveness relies too heavily on reflection, e.g., the constructor parameter names are required to find the expressive constructor, which can not be satisfied even by Java. Perhaps implementation of type expressiveness requires modification of current compilers, or the creation of a new compiler dedicated to implementing type expressiveness for each existing programming language.   
\end{itemize}

Not covered in this paper is the \emph{implied interface}, which is based on how an object is used.  For example, in C++ template programming, how an template object is used constitutes the implied interface of the template type, and there is no limit what actual type the template type could be.  Along this trend, Python allocated the implied interface on the object rather than the type of the object.  It actually demotes type into prototype, whose behaviour definition can be changed at runtime, and allows the prototype of an object to change at runtime; however, it also abandons the safety net of guaranteed behaviour before runtime.  The advantage of using implied interface is to reduce the need to change exiting code to abstract out a common explicit interface.  Perhaps one way to take advantage of implicit interface while still enjoying the safety net of guaranteed behaviour before runtime is to constrain the spec of an spec instance in a set, as shown in the following, in which a spec instance \texttt{mySpec} can be either a \texttt{Bool} or a \texttt{String} spec.
\begin{verbatim}
{Bool, String} mySpec;
\end{verbatim}

\acks

\bibliographystyle{abbrvnat}


\end{document}